\documentclass[modern]{aastex63}

\lefthead{Uzuo et al.}
\righthead{CO absorption in Circinus}

\newcommand{\bea}{\begin{eqnarray} }
\newcommand{\eea}{\end{eqnarray}}

\newcommand{\kms}{km s$^{-1}$ }
\begin{document}


\title{Circumnuclear Multi-phase Gas in Circinus Galaxy IV:  absorption owing to high-$J$ CO rotational transitions}
 
\author{%
Taisei Uzuo
}
\affiliation{Kagoshima University, Graduate School of Science and Engineering, Kagoshima 890-0065, Japan}

\correspondingauthor{Keiichi Wada}
\email{wada@astrophysics.jp}

\author{
Keiichi Wada
}%
\affiliation{Kagoshima University, Graduate School of Science and Engineering, Kagoshima 890-0065, Japan}
\affiliation{Ehime University, Research Center for Space and Cosmic Evolution, Matsuyama 790-8577, Japan}
\affiliation{Hokkaido University, Faculty of Science, Sapporo 060-0810, Japan}

\author{Takuma Izumi}%
\affiliation{National Astronomical Observatory of Japan, Mitaka, Tokyo 181-8588, Japan}
\affiliation{Department of Astronomical Science, The Graduate University for Advanced Studies, SOKENDAI, 2-21-1 Osawa, Mitaka, Tokyo 181-8588, Japan }
\affiliation{NAOJ Fellow}

\author{Shunsuke Baba}%
\affiliation{National Astronomical Observatory of Japan, Mitaka, Tokyo 181-8588, Japan}
\affiliation{JSPS Fellow}

\author{Kosei Matsumoto}%
\affiliation{ISAS/University of Tokyo}

\author{Yuki Kudoh}%
\affiliation{Kagoshima University, Graduate School of Science and Engineering, Kagoshima 890-0065, Japan}
\affiliation{National Astronomical Observatory of Japan, Mitaka 181-8588, Japan}
\affiliation{ALMA-J Grant Fellow}

\begin{abstract}

{We studied the absorption features of CO lines against the continuum originating from the heated dust in the obscuring tori around active galactic nuclei (AGNs). We investigated 
the formation of absorption lines corresponding to the CO rotational transitions using 
three-dimensional non-LTE line transfer simulations considering the
 dust thermal emission.}
As in Papers I--III of this series, we performed post-processed radiative transfer calculations using the ``radiation-driven fountain model'' \citep{wada2016}, which yields a geometrically thick obscuring structure
around the nucleus. This model is consistent with the spectral energy distribution of the nearest type-2 Seyfert galaxy, the Circinus galaxy.
{We found that the continuum-subtracted channel maps of $J = 4-3$ and higher transitions show 
absorption regions along the disk mid-plane for an edge-on viewing angle. 
The spectra consist of multiple absorption and emission features, reflecting the internal inhomogeneous and turbulent structure of the torus.
The deepest absorption feature is caused by the gas on the near-side of the torus between $r =10$ and 15 pc, which is located in front of the AGN-heated dust inside $r \simeq 5$ pc.
We also found that a spatial resolution of 0.5--1.0 pc is necessary to resolve
the absorption features. Moreover, the inclination angle must be close to the edge-on angle (i.e., $ \gtrsim 85^\circ$) to observe the absorption features. The findings of the present study imply that combining our radiation-hydrodynamic model with high-resolution observations of CO (7-6) by 
ALMA can provide new information about the internal structure of the molecular tori in nearby AGNs}.

\end{abstract}

\keywords{galaxies: active -- galaxies: nuclei -- galaxies: ISM -- radio lines: ISM -- radiative transfer} 

%
\section{INTRODUCTION}
%
In the standard picture of active galactic nuclei (AGNs), the broad emission line region is 
enshrouded by an optically as well as geometrically thick material \citep{antonucci93}.
However, the actual structure of the obscuring material is not well understood; hence, it is often postulated to be a torus-like object \citep{urry95} \citep[see also][for review] {netzer2015}.
This situation is rapidly changing 
owing to the high spatial resolution of the Atacama Large Millimeter/submillimeter Array (ALMA). Recently, the obscuring material consisting of molecular gas and dust 
has been resolved in several nearby AGNs \citep{garciaburillo2016, izumi2018, combes2019, imanishi2016, imanishi2020}.

{However, the internal structure of the molecular tori revealed by ALMA is still unclear; for example, it is unclear whether the tori are clumpy or smooth, and the internal turbulent structure, even in nearby AGNs, is unknown.
A promising approach to answer these questions is to make a comparison between
synthetic observations based on realistic
theoretical models and actual multi-wavelength observations of the tori.
Such a comparison is especially important for observations conducted with a resolution of $\sim 0.01"$, which 
corresponds to sub-pc scales in nearby AGNs. However, in such comparisons, classical torus models, which phenomenologically assume
the geometry, size, and internal structure of the torus, cannot  reveal much information regarding the obscuring material in AGNs.
}

In contrast to phenomenological torus models, \citet{wada2012} proposed 
a physics-motivated picture of the obscuring structures around the central engine in 
AGNs based on three-dimensional (3D) radiation-hydrodynamic models.
Their ``radiation-driven fountain'' model, in which both
outflowing and inflowing gases are driven by 
the radiation from the accretion disk, naturally forms 
a geometrically thick disk extending for tens of parsecs \citep[see also][]{namekata2016,dorod2016, dorod2017,chankrolik2017,williamson2020}.
This quasi-steady circulation of gas obscures the central source; hence, 
  the differences in the spectral energy distributions (SEDs) 
 of typical type 1 and type-2 Seyfert galaxies
 can be explained by invoking different viewing angles \citep{schartmann2014}.
In \citet{wada2016} (hereafter, W16), we applied this radiation-driven fountain model to 
the Circinus galaxy, which is the nearest ($D=$ 4 Mpc) type 2 Seyfert galaxy.
Post-processed three-dimensional radiation transfer calculations were used to model the dust thermal emission; based on this,  
``polar'' emission in the mid-infrared band (12 $\mu {\rm m}$) was detected, which is often observed in nearby Seyfert galaxies \citep[][]{hoenig2013,tristram2014, asmus2016}.
In addition, it was confirmed that the viewing angle should be larger than 75$^\circ$ (i.e., close to edge-on) to
explain the observed SED and the 10 $\mu$m absorption feature of the Circinus galaxy \citep{prieto2010}.

{
To confirm whether the radiation-driven fountain model can
explain observational properties in addition to the SED
in the Circinus galaxy,
we recently conducted a quantitative comparison between this model and multi-wavelength observations.}
In \citet{wada2018a} (hereafter, Paper I), we calculated molecular line emissions, such as those from $^{12}$CO (hereafter, CO)(2--1), 
using the best-fit model of W16. The model results were further compared with ALMA Cycle-4 observations of {rotational lines of CO and of the fine structure line of atomic carbon, [CI] 1-0} \citep{izumi2018} (hereafter, Paper II).
The same model was also used to calculate optical emission lines from ionized gas, such as [OIII] 5007 \AA. We found that 
radiation-driven outflows irradiated by the central source can help explain the properties of the observed emission lines (e.g., the {Baldwin, Phillips \& Terlevich (BPT) diagrams \citep{baldwin1981}})
and the structure of the narrow-line regions (NLRs) \citep{wada2018b} (hereafter, Paper III).
{In addition, the X-ray spectral properties were investigated 
using the same strategy, and they were found to be consistent with
observations (Buchner et al. 2021; Ogawa et al. in prep.).}

{
In Papers I and II, we focused on the molecular and atomic \textit{emission} lines.
However, if the background source is sufficiently warm, then the physical conditions of the cold ISM located between the observer and the source can be derived
using \textit{absorption} lines. 
Low-$J$ CO absorption has been detected in some radio galaxies against the emission from
radio jets
\citep{israel1990, israel1991, jaffe1994, espada2010, kameno2020}. In ultra-luminous infrared galaxies (ULIRGs), CO absorption owing to rotation-vibration transitions has also been detected against the thermal emission from heated dust in the near-infrared \citep{spoon2004, spoon2005, shirahata2013, shirahata2017, baba2018}. }

{
However, CO absorption features owing to rotational transitions have not been detected in radio-quiet AGNs or Seyfert galaxies at submillimeter wavelengths \citep[for example,][]{okuda2013}.   
This is in contrast to the observations of the CO absorption in ULIRGs owing to the rotation-vibration transitions \citep{shirahata2013, shirahata2017}.
Therefore, it is important to understand the reason for the absence of CO absorption lines in low-luminosity AGNs
and the conditions necessary to form CO absorption features  owing to the rotational transitions against 
the dust continuum.  We suspect that the {cool} dust ($<$ 100 K) in the tori of low-luminosity AGNs may serve as the background source at submillimeter wavelengths.
{There are several possible reasons 
for the absence of CO absorption features;
1) it could be due to the low-luminosity dust continuum in 
AGNs,  2) there is no temperature gradient between the dust and gas, 
3) the covering fraction of the foreground gas is small compared to
the extent of the background source, and 4)
the opacity of the foreground molecular gas is low.
Alternatively, it is possible that the spatial resolution of past observations in the millimeter and submillimeter wavelength range was not sufficiently high\footnote{In \citet{okuda2013}, 
the spatial resolution was approximately $2''.1-2''.7$ with Nobeyama Radio Observatory (NRO)/RAINBOW interferometer   and $0''.73-1''.2$ with the IRAM Plateau de Bure Interferometer.}
compared to the size of 
the background source and molecular tori. In this case, the extended emission line features may contaminate the
absorption features of smaller size.
Therefore, it is important to study the effect of 
spatial resolution on the observed line properties, especially with the advent of ALMA.
}

{To clarify the aforementioned points theoretically, 
we need realistic 3D models 
that include the dust component and molecular gas distributions,
  from which we can calculate both the background dust continuum 
  and molecular line intensities.
  Note that the temperature of the dust associated with the ISM should be self-consistently determined 
  from the radiation field of the AGN.
  Therefore, we use the radiation-driven fountain model, as in the previous papers of this series.
However,
in the theoretical calculations presented in Papers I and II, 
 we did not consider the dust emission.
 Therefore, in this work,
we performed new 3D radiative transfer calculations 
including both the continuum emission owing to the AGN-heated dust 
and the rotational transition lines of CO.
}

{
In this study, CO lines were calculated using the same non-local thermodynamic equilibrium radiative transfer code used in Papers I and II. 
The background dust thermal emission was calculated using the 3D open source radiative transfer code } \verb+RADMC-3D+ \citep{dullemond2012}. 
 {In contrast to Papers I and II, in which relatively low-$J$ transitions were considered, 
 in the present study, we investigated transitions up to $J = 9-8$. This is because we expect 
 the thermal dust emission to be brighter for higher-$J$ transitions, which can be observed with ALMA.}

 {In this fourth paper of the series, we would like to address the following questions:  
(1) Can the absorption features be observed at millimeter and submillimeter wavelengths against 
the continuum originating from the AGN-heated dust of several tens of kelvin in the radiation-driven fountain model? 
(2) What are the conditions necessary to realize point (1) (e.g., the transitions suitable for observations, spatial resolution of the synthetic observations, and viewing inclination angles)?
(3) How do the absorption features correspond to 
the internal spatial and kinematic structures of the obscuring material?
The answers to these questions
would be useful for conducting future high-resolution observations with ALMA, for example, using {CO $(J=7-6)$ (806.7 GHz) for local AGNs} \footnote{
In this study, we focused only on the CO rotational transitions. The radiative transfer calculations for the CO rotation-vibration transitions in the near-infrared region
will be discussed in a subsequent paper (Matsumoto et al., in prep.).}.}

{One may wonder that if the internal density of the tori is clumpy, we could also use high-density tracers, such as HCO$^+$.
In fact, \citet{yamada07} studied HCO$^+$ and HCN line structures for 
a hydrodynamic model of the starburst-driven torus \citep{wada02}; their method is basically the same as the one used in 
this paper,  but the AGN feedback was not included. Because CO is abundant and the background dust continuum is brighter in sub-millimetre,  we here focus on high $J$ lines of CO (the critical density for CO (7-6) is $\sim 10^5$ cm$^{-3}$). 
Absorption for HCO$^+$ or other lines will be discussed elsewhere. }

The remainder of this paper is organized as follows. In \S 2, our input model, namely the radiation-driven fountain model, is briefly explained, and the radiation transfer calculations for the dust and CO lines are described. 
The results, including the CO intensity maps and spectra, are presented in \S 3. 
The implications for future observations are discussed in \S 4, and finally, the paper is concluded in \S 5.

%
\section{NUMERICAL METHODS AND MODEL}
%

\subsection{Input model: Radiation-driven fountain}

The input model used for the line transfer calculations in this work is 
the same as that used in Papers I--III.
This model consistently explains the infrared SED of the Circinus galaxy when 
the viewing angle is 75$^\circ$ or higher (W16).

\begin{figure}[h]
\centering
\includegraphics[width = 14cm]{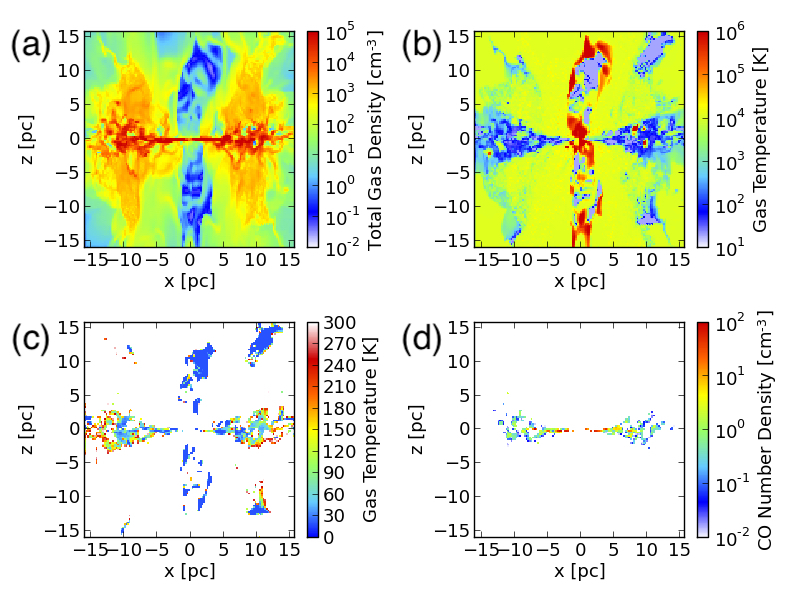}    
\caption{
Distribution quantities on $x-z$ planes of the radiation-driven fountain model used for the radiative transfer calculations.
(a) Total gas density (cm$^{-3}$).  (b) Gas temperature (K).  (c) Cold gas temperature (K). (d) CO number density (cm$^{-3}$).}
\label{wada_fig: 2}
\end{figure}

\begin{figure}[h]
\centering
\includegraphics[width = 14cm]{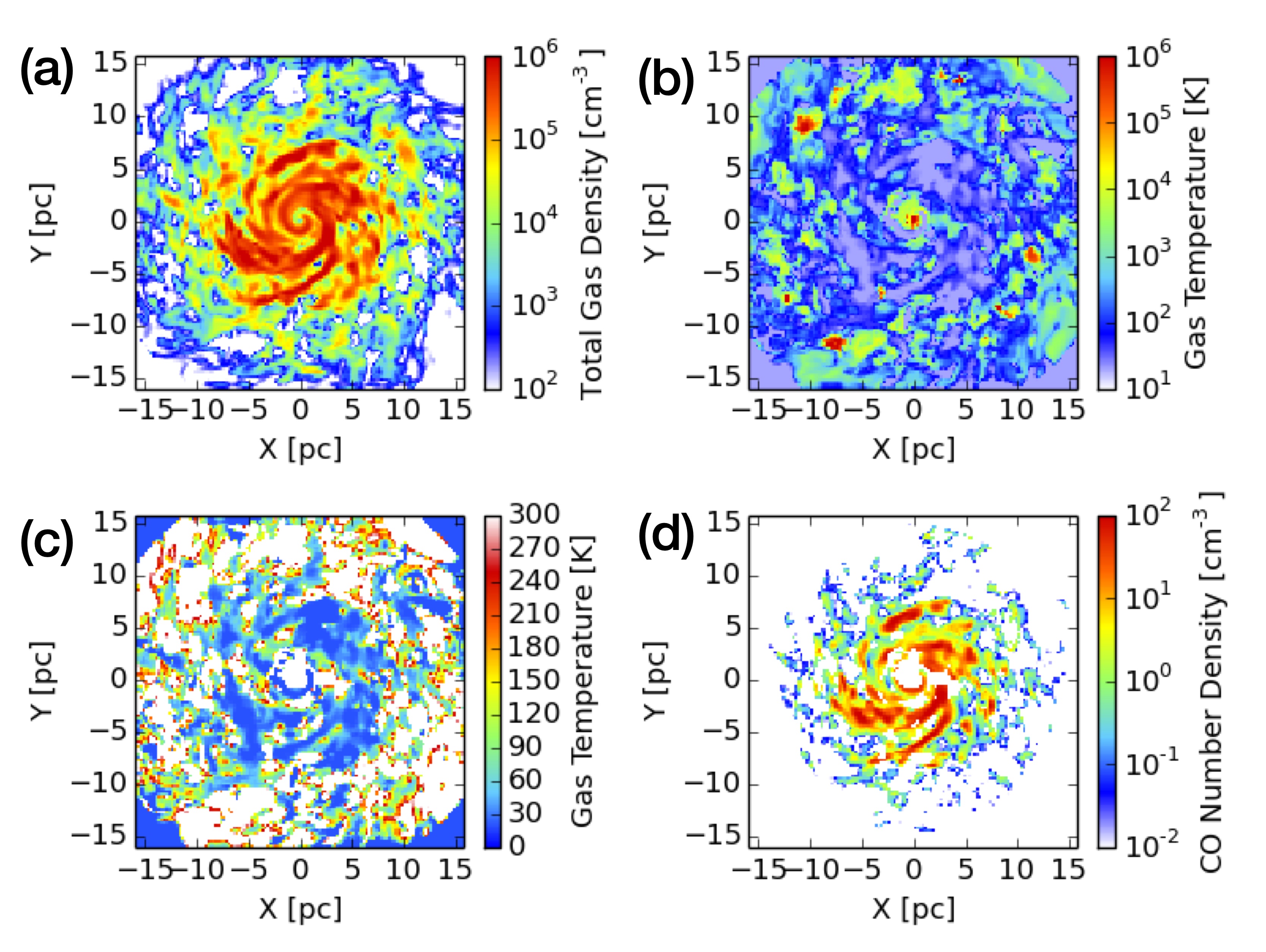}    
\caption{Same as Fig. \ref{wada_fig: 2}, but for 
face-on views.}
\label{wada_fig: map2}
\end{figure}

\begin{figure}[h]
\centering
\includegraphics[width = 8cm]{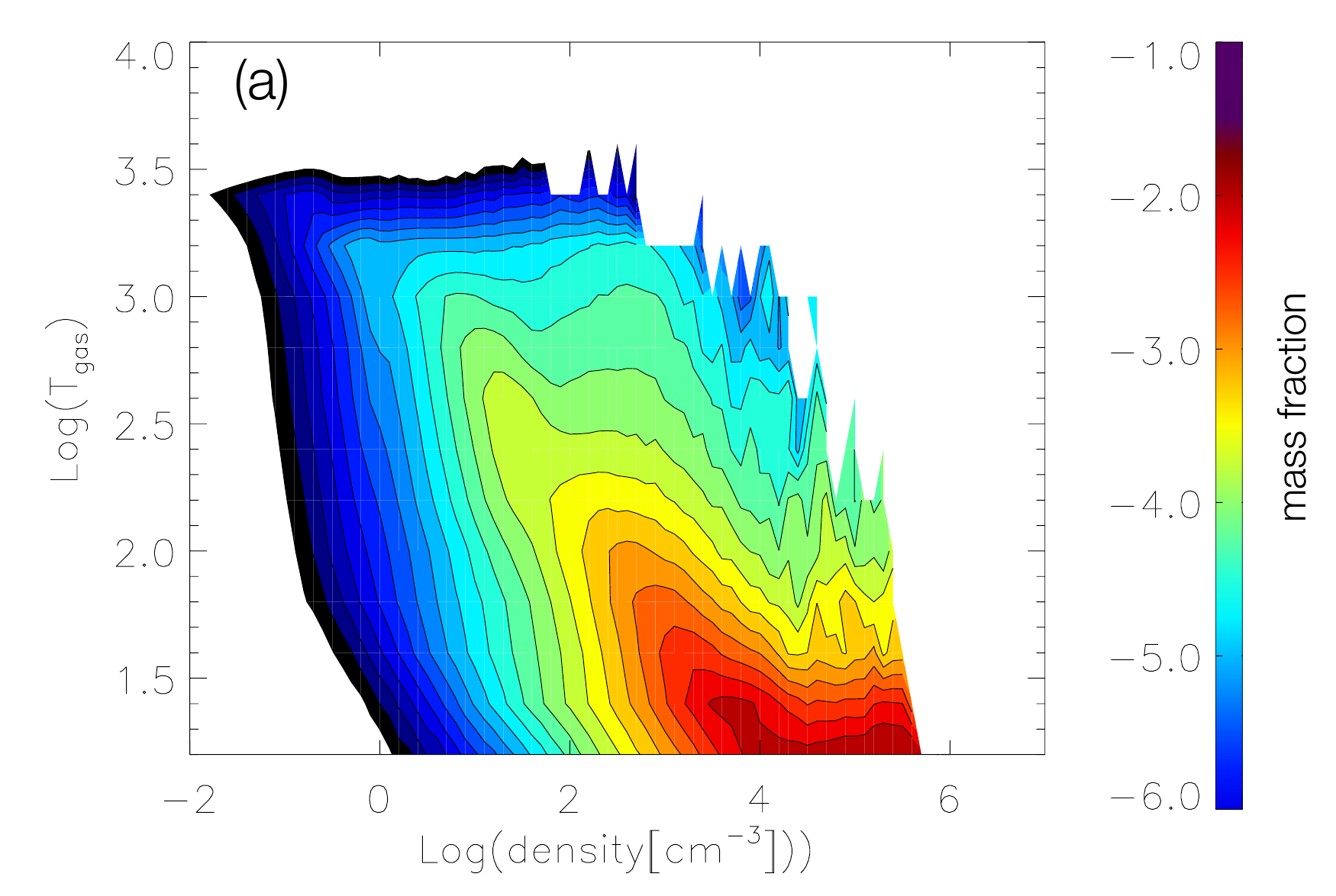}  
\includegraphics[width = 7.5cm]{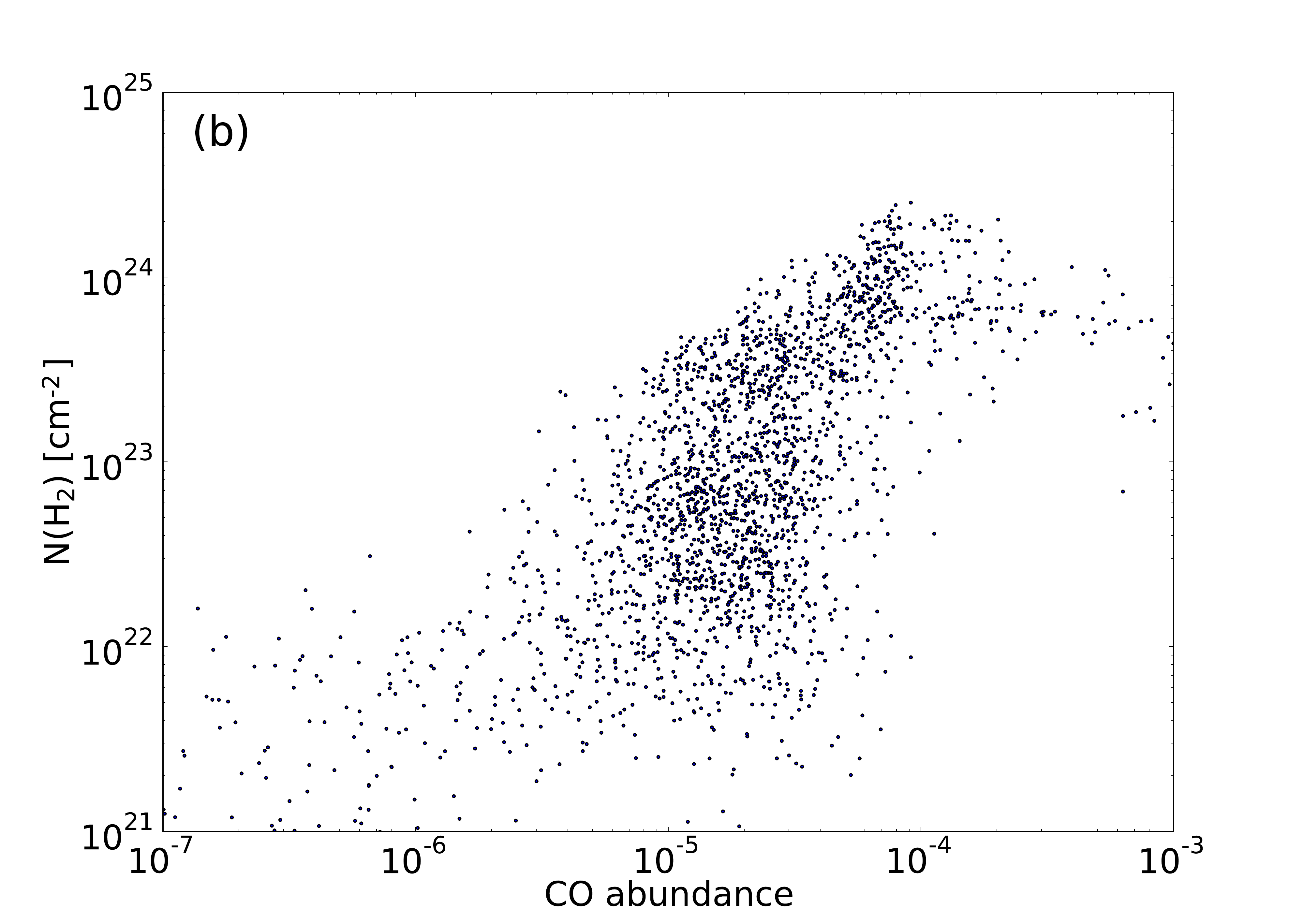}  
\caption{{(a) Phase diagram of the input data. (b) CO abundance distribution and the column density of H$_2$
in each grid cell (see also W18).}}
\label{wada_fig: phase}
\end{figure}


{
The hydrodynamic model was obtained using a 3D Eulerian hydrodynamic code \citep{wada2012, wada2015} with a uniform grid, which accounts for radiative feedback processes from the AGN using a ray-tracing method.
We assume a time-independent external potential:
$\Phi_{\rm ext}(r) \equiv -(27/4)^{1/2}[v_1^2/(r^2+a_1^2)^{1/2}+v_2^2/(r^2+ a_2^2)^{1/2}]$, where $a_1 = 100$ pc, $a_2 =
2.5$ kpc, $v_1 = 147$ km s$^{-1}$, and $v_2 = 147$ km s$^{-1}$ to represent a stellar potential.
We adopt implicit time integration for the radiative cooling.
To prepare quasi-steady initial conditions without
the radiative feedback from the AGN, we first evolve an axisymmetric and rotationally supported thin
disk with uniform density profiles with  1\% of random fluctuation in density. 
The radius of the gas disk in this case is approximately 16 pc, and the total gas mass was also assumed to be $2\times 10^6 M_\odot$.
After the thin disks is dynamically settled, the radiation 
feedback from the AGN and supernovae (SNe) feedback (the supernova rate is fixed to be 0.014 yr$^{-1}$
during the calculation) are turned on.}

We assumed solar metallicity and cooling functions for 20 K $\le T_{gas} \le 10^{8}$ K \citep{meijerink05, wada09}.
The central black hole mass was assumed to be $2\times 10^6 M_\odot$,
which is comparable to the value estimated from maser observations of Circinus \citep{greenhill2003}. 
The Eddington ratio was set to 0.2 and the bolometric luminosity to $L_{bol}=5 \times 10^{43}$ erg s$^{-1}$.
{
The ultraviolet flux is assumed to be $F_{UV}(\theta) \propto \cos \theta (1+2\cos \theta)$, 
where $\theta$ denotes the angle from the rotational axis ($z$-axis).
The X-ray radiation, on the other hand, assumed to be spherically symmetric  \citep{netzer1987}.
The UV and X-ray fluxes are calculated from the bolometric luminosity \citep{marconi2004, hopkins2007}. 
The total X-ray luminosity (2--10 keV) is $L_X = 2.8\times 10^{42} $ erg s$^{-1}$.
}


{Figures \ref{wada_fig: 2} and \ref{wada_fig: map2} shows the distributions of gas density, temperature, and CO number density in the 
input model.  As shown in the phase-diagram (Fig. \ref{wada_fig: phase}(a) ), 
the high density gas ($ > 10^4 $ cm$^{-3}$) is mostly cold ($T_g < 100$ K).  CO is more abundant in 
this high density region (Fig. \ref{wada_fig: phase}(b)) that forms a thin disk with spiral-like features (Fig.  \ref{wada_fig: map2}(d)).
}

{We solved the non-equilibrium X-ray dominated region (XDR) chemistry \citep{maloney96,
meijerink05} for all the 256$^3$ zones (i.e., a resolution of 0.125 pc).  At every hydrodynamic time step, 
the gas density, gas and dust temperatures, ionization parameters in the 256$^3$ grid cells, and the time step (typically 50-100 years)
are passed to the chemistry module. The chemistry module iteratively calculates
the chemical network to reach an equilibrium for the time step, and it  
returns the abundances of the species to the hydrodynamic part.
The abundance distributions are advected based on the gas velocity obtained in the
hydrodynamic part.
We used a selection of reactions from the chemical network described by \citet{meijerink05, adamkovics2011}
for 26 species:  H, H$_2$, H$^+$,  H$_2^+$, H$_3^+$, H$^-$, e$^-$, O, O$_2$, O$^+$,  O$_2^+$, O$_2$, H$^+$, OH,
OH$^+$, H$_2$O, H$_2$O$^+$, H$_3$O$^+$, C, C$^+$, CO, Na, Na$^+$, He, He$^+$, and HCO$^+$
 \footnote{To save computational memory, we solve the advection of the species between the grid cell only for
H, H$^+$, H$_2$, O, O$^+$, H$_2$O, OH, C, C$^+$, and $e^-$}. 
}

{
The size of the chemical network is the maximum one to 
solve the non-equilibrium chemistry in the $256^3$ hydrodynamic grid cells for more than 10000 time steps, but 
it ensures that the the important species for the present study, such as CO, is 
consistent with the results with 1-D calculations based on the full network. 
For example, for $N_{\rm H} = 10^{24}$ ($10^{22}$) cm$^{-2}$ , $n_{\rm H} = 10^{5.5}$ cm$^{-3}$ and 
the X-ray flux $F_X = 160$ erg cm$^{-2}$ s$^{-1}$, the full network including S- and N-bearing species, \citet{meijerink05} predicted that the CO abundance is $\sim 10^{-4}$ ($5\times 10^{-6}$) relative to H$_2$, whereas it is $10^{-4}$ ($2\times 10^{-6}$)  in the high density region ($n_{\rm H} > 10^5$ cm$^{-3}$) in our model (Fig. \ref{wada_fig: phase}(b)). Moreover, the observed CO (3-2) peak flux of the molecular torus of the Circinus galaxy obtained by ALMA (Paper II)
is consistent with our prediction (see Appendix, Fig. \ref{wada_fig: appendix}) within a factor of two (i.e., 16 mJy/beam vs. 28 mJy/beam)\footnote{More generally, in
a comparison study among various PDR chemical codes including the XDR model of \citet{meijerink05},  \citet{roellig2007}
pointed out that \textit{a small network can give similar results as a big network}. 
}}. 
}

{The metallicity could also affect the abundance of not only for CO, but also that of other species.  
In Paper III, we calculated the optical emission lines of the ionized outflows, assuming the solar metallicity.
We found that the emission lines from the ISM around the AGN
are consistent with the observed properties, for example such as the line ratios, e.g.,  [OIII]/H$\beta$ and [SII]/H$\alpha$.
Therefore, the assumption of the solar metallicity in this paper is a reasonable first choice for the AGN model.
}

\subsection{Radiative transfer of CO lines}
We used a 3D line transfer code \citep{wada05,yamada07}, which is based on a Monte Carlo and long-characteristic radiative transfer code \citep{hoge00}.  
The rate equations and radiative transfer equations were iteratively solved,
considering that photon packages propagate into each grid cell. We solved the following equation 
\begin{eqnarray}
\frac{d I_\nu}{d \tau_\nu} = - I_\nu + S_\nu,
\label{eq: 1}
\end{eqnarray}

where the source function $S_\nu$ is given by
\begin{eqnarray}
S_\nu =   \frac{j_{{\rm CO},\nu}
 + \alpha_{d,\nu} B_{\nu} (T_d)}
 {\alpha_{{\rm CO}, \nu} + \alpha_{d, \nu}}. 
 \label{eq: 2}
 \end{eqnarray}
The optical depth is
defined as $d \tau_\nu \equiv (\alpha_{{\rm CO}, \nu} + \kappa_{d, \nu} \rho_d) ds$,
{where {the dust-to-gas mass ratio is fixed as an assumption throughout the paper, }i.e., $\rho_d = 0.01 \rho_g$ is the dust density and $\rho_g$ is the gas density (determined using the hydrodynamic model) in each grid cell.}
The dust temperature $T_d$ in each grid cell is computed using \verb+RADMC-3D+ (\S 2.3).
 
The coefficients $j^{ul}_\nu = j_{{\rm CO}, \nu}$ and $\alpha^{ul}_\nu = \alpha_{{\rm CO}, \nu}$ for a spectral line can be determined using
the absorption and emission between levels $u$ and $l$ with number density
$n_u$ and $n_l$, respectively, such that

\bea
j_\nu^{ul} &=& \frac{h\nu_0}{4 \pi} n_u A_{ul} \phi(\nu), \label{eq: 3}\\
\alpha_\nu^{ul} &=& \frac{h\nu_0}{4 \pi} (n_lB_{lu} - n_u B_{ul}) \phi(\nu), \label{eq: 4}
\eea
where $\phi(\nu)$ is a line-profile function that peaks around the frequency $\nu_0 = (E_u -E_l)/h$, and $A_{ul},B_{lu}$ and $B_{ul}$ are the Einstein probability coefficients.
The line profile $\phi(\nu)$ is given by
\bea
\phi(\nu) = \frac{1}{\sqrt{\pi}\sigma\nu_0} \exp\left[-\frac{c^2 (\nu-\nu_0)^2}{\nu_0^2 \sigma^2}\right].
\label{eq: 5}
\eea
We assumed that the width $\sigma$ of the line profile is influenced by micro-turbulence, that is, a hypothetical turbulent motion inside each grid cell, and used the turbulent velocity $v_{turb} = \sigma$ as a free parameter (see \S 4 for details).
  In the fiducial model, we assumed $v_{turb} = 5$ \kms.
  {Note that the line width of molecular lines observed in external galaxies is usually
   much larger than the value assumed here.
  This is because the velocity shear owing to the galactic rotation may be convolved with the observed line width if the beam size is not sufficiently small. Additionally, $v_{turb}$ here corresponds to the velocity dispersion in a grid cell (0.25 pc). }
  
The mean intensity was evaluated using the Monte Carlo approach \citep{hoge00}, 
in which the integration is performed along the rays incident on each cell from 
infinity. 

The 3D hydrodynamic grid data (i.e., density, temperature, abundances, and three components of the velocity) from the 256$^3$ grid cells
were averaged to produce data corresponding to 128$^3$ grid cells (i.e., having a spatial resolution of 0.25 pc) to 
reduce the computational cost. These data were fed into the 3D non-LTE line transfer code, as described in the next section, to
derive the line intensities.
For each grid cell, $1500$ rays were considered. We calculated {15 rotational transitions of CO (i.e., $J=1-0, 2-1, ... 15-14$)}.
The level populations converged 
with an error less than $10^{-3}$ for $J \le 10 $ after 10 iterations. 

Once the radiation field and optical depth were determined for all grid cells, we ``observed'' it from 
arbitrary directions and generated 3D data cubes (of positions and line-of-sight velocity) for selected transitions.

\subsection{Dust continuum}

\begin{figure}[h]
\centering
\includegraphics[width = 7cm]{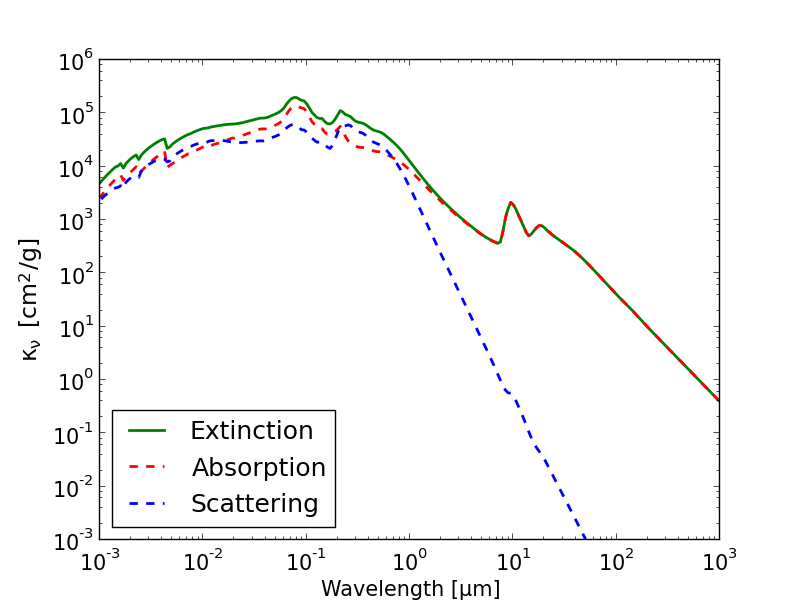} 
\caption{Opacity curve adopted for the RADMC-3D calculations. The absorption and scattering components are shown by the red and blue dashed lines, respectively.}
\label{wada_fig: 1}
\end{figure}

To obtain the source function for the dust continuum in each grid cell, 
we used the 3D radiative transfer code \verb+RADMC-3D+ (version 0.41) \citep{dullemond2012}\footnote{https://www.ita.uni-heidelberg.de/~dullemond/software/radmc-3d/}. 
The SED of the AGN was taken from \citet{schartmann2005}. 
The opacity curve used is shown in Fig. \ref{wada_fig: 1}.
The size of the
dust grains was assumed to be between 0.01$\mu$m and 0.1$\mu$m, and their distribution function was adopted from \citet{mrn1977}.  
{We assumed that the dust is composed of silicate and graphite with a mass ratio of 0.625:0.375 \citep{mrn1977, schartmann2005}.
\citet{schartmann2014} calculated the infrared spectrum in the Circinus assuming the same dust model, and 
found that the SED and the 10$\mu$m absorption feature well fit the observations, if the inclination angle is larger than 
75$^\circ$.  
}

\begin{figure}[h]
\centering 
\includegraphics[width = 5cm]{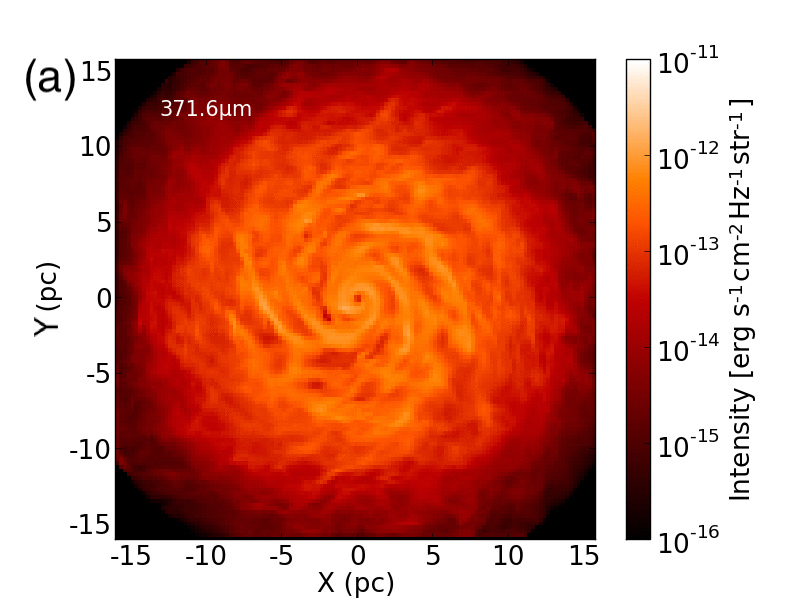}
\includegraphics[width = 5cm]{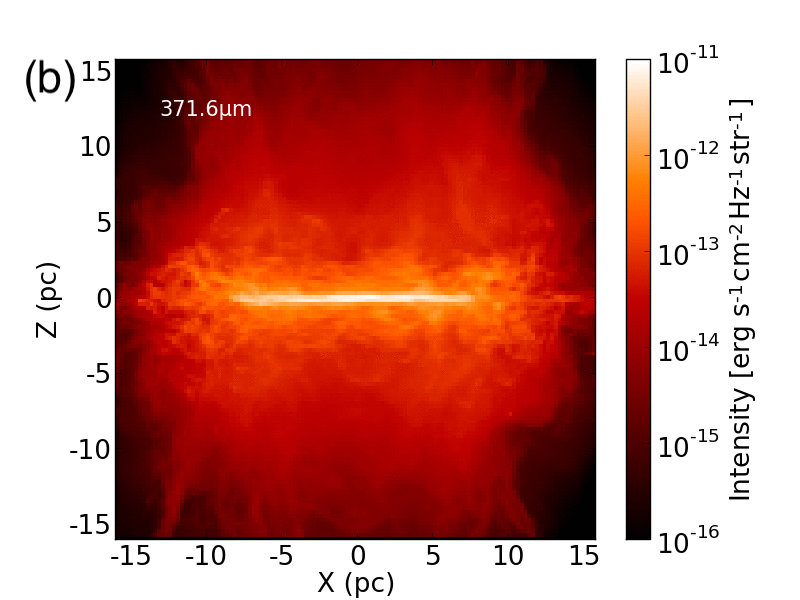}
\includegraphics[width = 5cm]{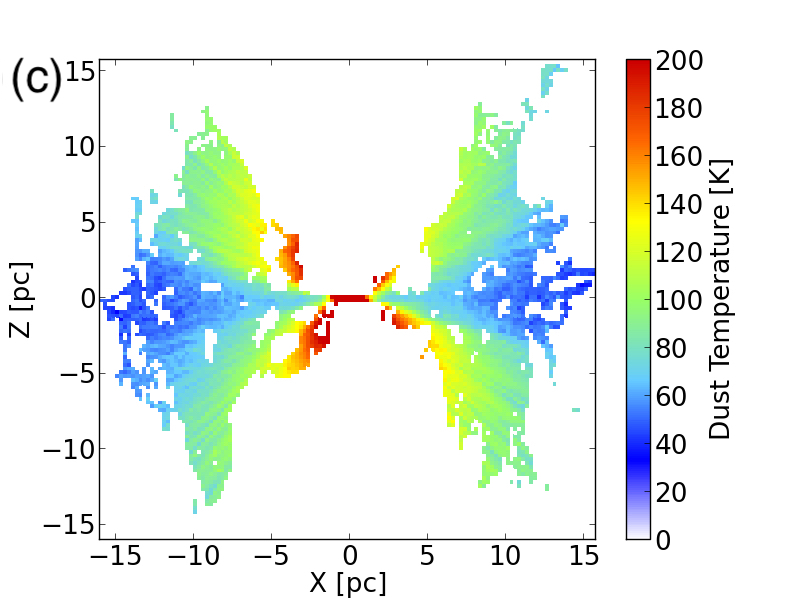}
\caption{
(a) Face-on view of the thermal dust emission ($B_\nu(T_d)$ at $\lambda = 371.6 \mu {\rm m}$) corresponding to CO(7--6). (b) Same as (a) but for edge-on view. (c) Dust temperature $T_d$ on the x-z plane. Note that only the regions with $\rho_g > 10^{21}$ g cm$^{-3}$ are plotted.
 }
\label{wada_fig: 2b}
\end{figure}

The 2D distribution of the dust temperature $T_d$ and the continuum dust emission $B_\nu (T_d)$ given by Eq. (\ref{eq: 2}) are shown in 
Fig. \ref{wada_fig: 2b}. The size of each grid cell is 0.25 pc in this case.
It is clear that most of the dust emission is concentrated in the disk mid-plane (i.e., $|z| \lesssim 3$ pc), and it forms a spiral-like structure in the face-on view.

%
%
\section{RESULTS}
%
\subsection{Intensity maps and spectra of CO rotational transition lines}

\begin{figure}[h]
\centering
\includegraphics[width = 6cm]{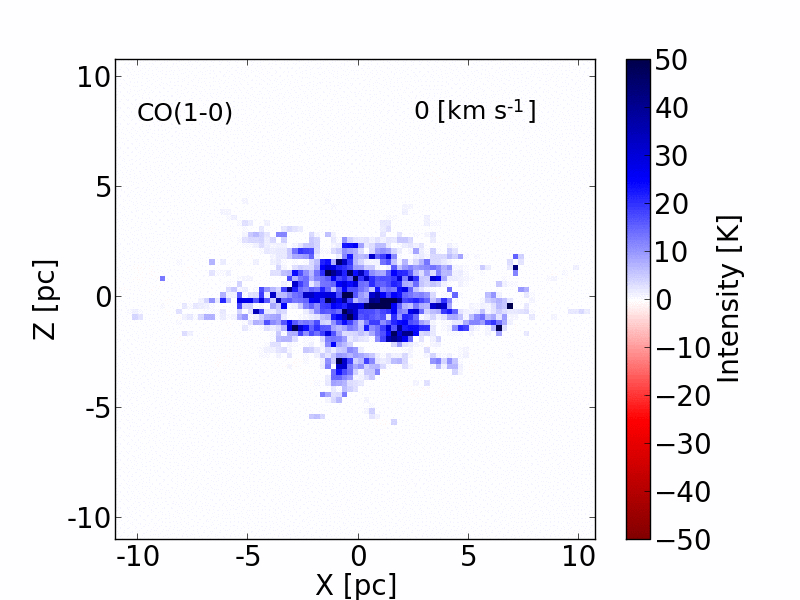}
\includegraphics[width = 6cm]{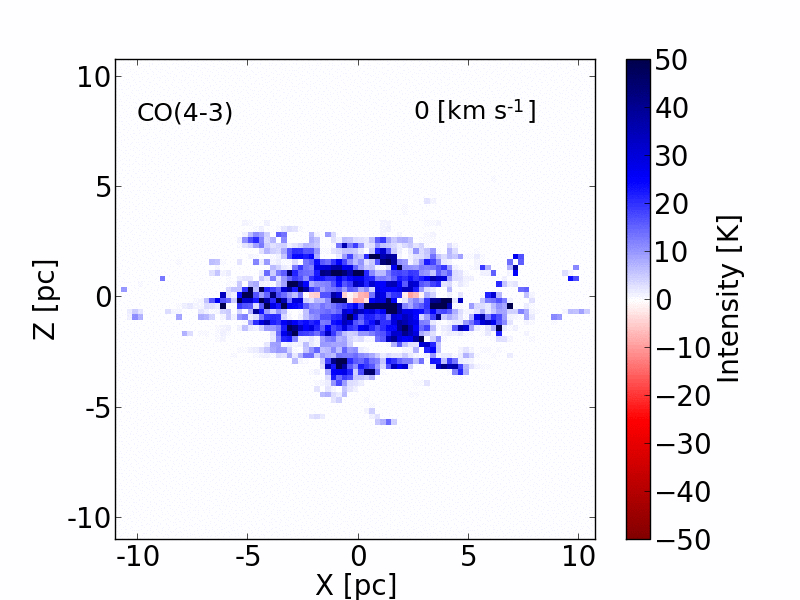} \\
\includegraphics[width = 6cm]{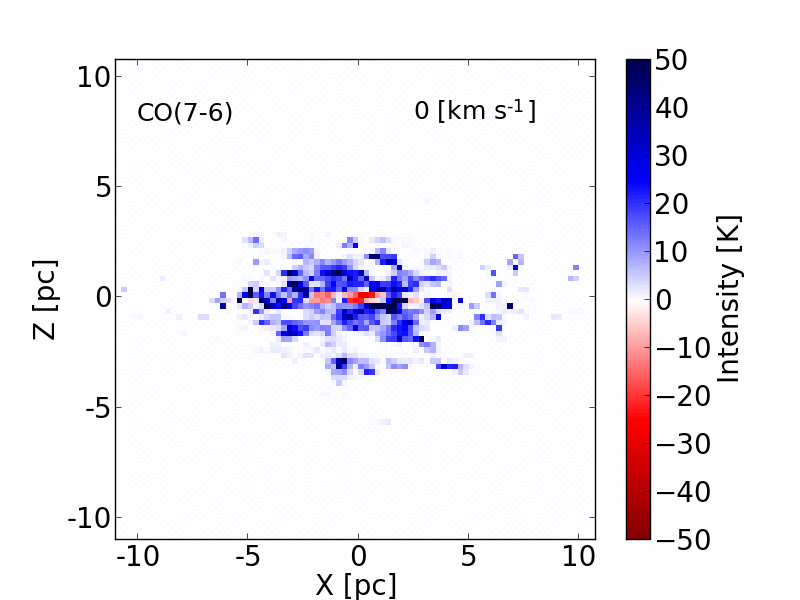}
\includegraphics[width = 6cm]{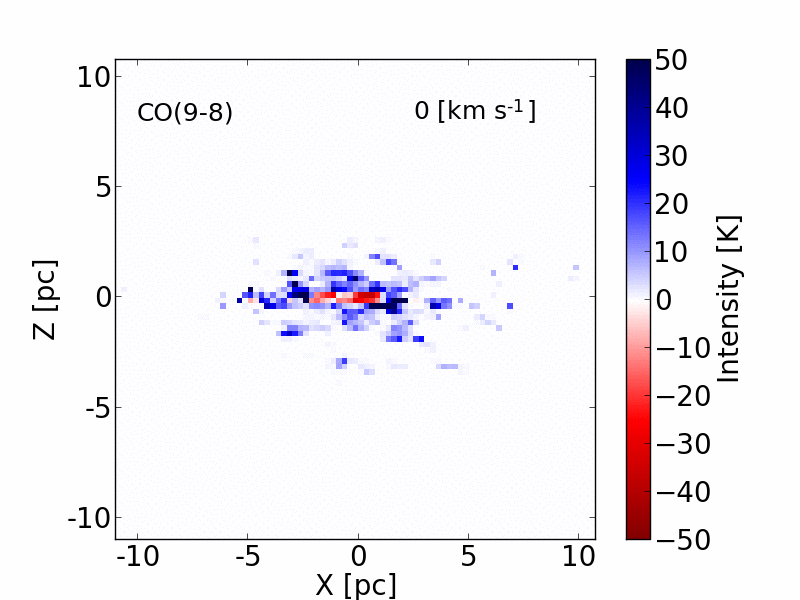}

\caption{Channel maps for $v=0$ \kms. The colorbar represents the intensity after subtracting the dust continuum. The panels represent CO(1-0), (4-3), (7-6), and (9-8) intensity maps for an inclination angle $i = 90^\circ$ (edge-on view) and azimuthal angle $\phi_{obs} = 50^\circ$.  
$\phi_{obs}$ is defined anticlockwise from the $x$-axis for $x > 0$.} 
\label{wada_fig: 3}
\end{figure}

Figure \ref{wada_fig: 3} shows the intensity maps for $v = 0 $ \kms at 
115.3, 461.0, 806.7, and 1036.9 GHz, corresponding to the rest frame frequencies of CO(1--0), CO(4--3), CO(7--6), and CO(9--8), respectively,  
for an inclination angle $i = 90^\circ$ (i.e., edge-on). We assumed $v_{turb} = 5$ km s$^{-1}$ (the effect of $v_{turb}$ 
will be discussed in \S 4).
In these maps, the dust continuum was subtracted from the total intensity, and hence, the negative intensity (i.e., red regions)
represents absorption.
The CO(1--0) map shows that the emission dominates the entire region, forming a thick disk with a scale height of $\sim 3 $ pc.
In contrast, absorption regions can be seen in the CO(7--6) and CO(9--8) maps, which appear as thin disks ($|z| < 1$ pc).
This is because the thermal emission from the AGN-heated dust 
as well as the distribution of CO 
are concentrated along the thin disk, as seen in Fig. \ref{wada_fig: 2b}.

\begin{figure}[h]
\centering
\includegraphics[width = 7.cm]{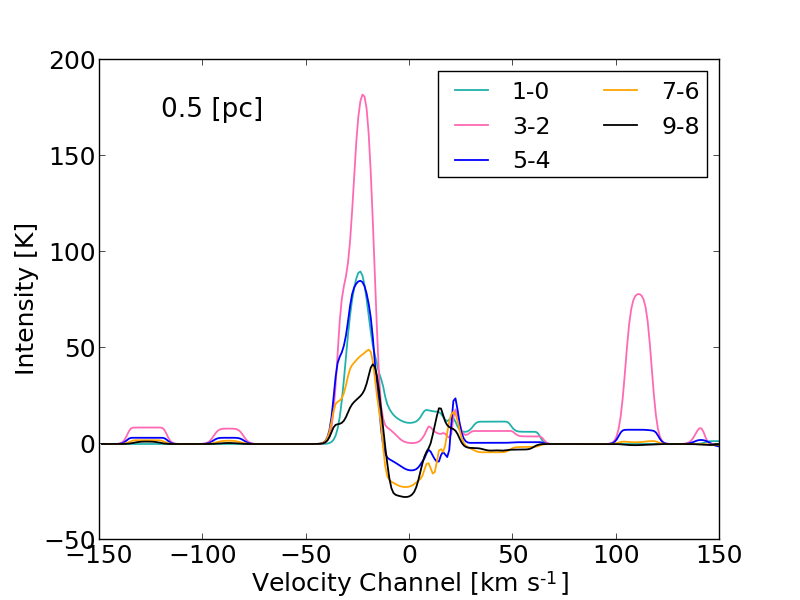}
\includegraphics[width = 7.cm]{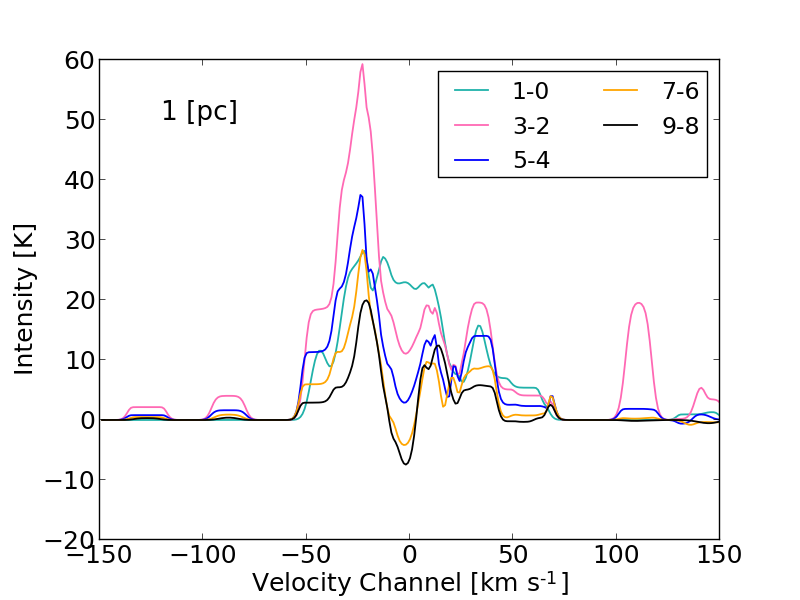}
\includegraphics[width = 7.cm]{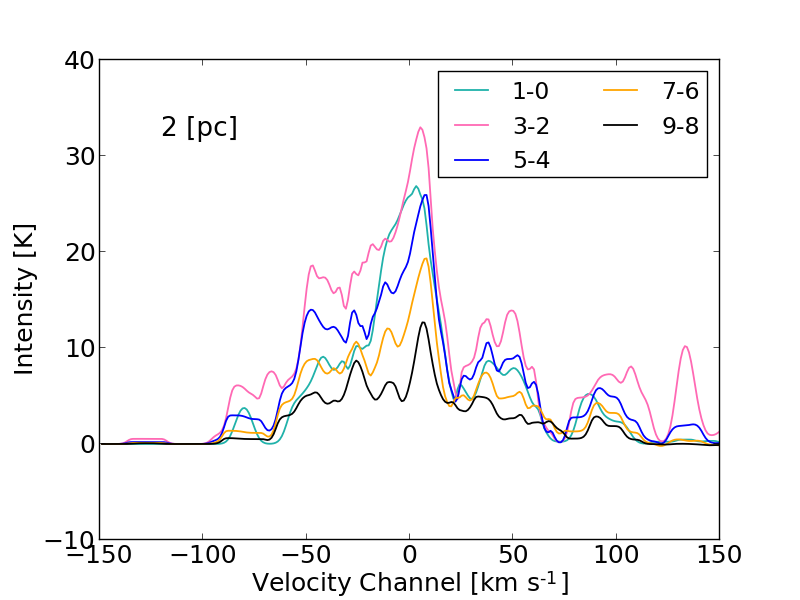}
\includegraphics[width = 7.cm]{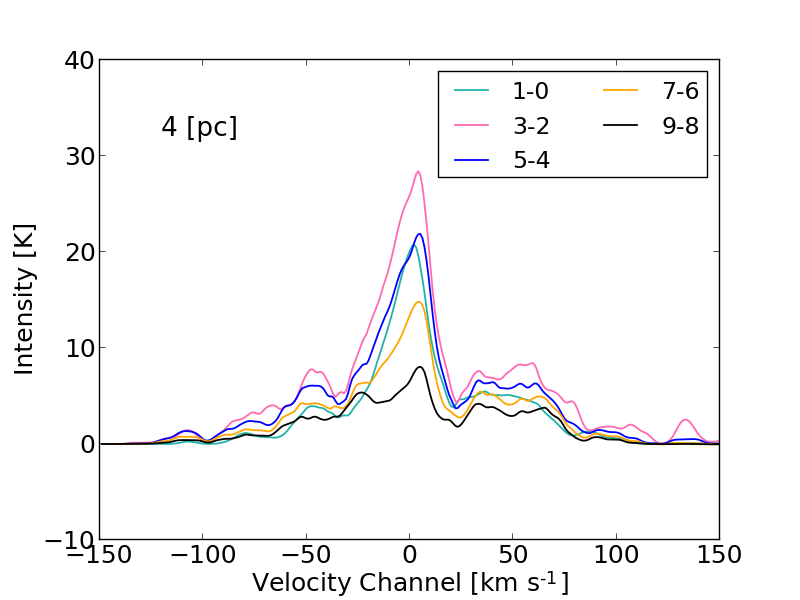}
\caption{CO(1--0), (3--2), (5--4), (7--6), and (9--8) spectra toward the center with the dust continuum subtracted, for $i = 90^\circ$, $\phi_{obs} = 50^\circ$, and $v_{turb} = 5$ \kms.
The beam sizes are 0.5, 1, 2, and 4 pc.
Note that the range of intensity is different in each plot.
See Appendix for the spectra before subtracting the continuum.}
\label{wada_fig: 5}
\end{figure}

{Note that if the observed beam size is not as small as the
scale height of the absorption regions in Fig. \ref{wada_fig: 3}, then the emission may contaminate the absorption, and the absorption features may not be observed.
Thus, to evaluate this phenomenon quantitatively, 
we studied the effect of different averaging areas on the spectra. 
Figure \ref{wada_fig: 5} shows {five} {continuum-subtracted} spectra\footnote{The raw spectra
before subtracting the dust continuum are presented in the Appendix.}
 (CO(1--0) to CO(9--8)) toward the center
for four different averaging areas:  0.5 pc $\times $ 0.5 pc, 1 pc $\times $ 1 pc, 2 pc $\times $ 2 pc, and 
4 pc $\times $ 4 pc. We assumed that the viewing angle is edge-on.
The four plots in Fig. \ref{wada_fig: 5} roughly correspond to observations with four different ``beam" sizes.}
For higher-$J$ transitions, the absorption features around 0 \kms are more prominent for beam sizes of 0.5 and 1.0 pc.
However, for beam sizes of 2 and 4 pc, only the emission lines with 
self-absorption-like features are observed.

\begin{figure}[h]
\centering
\includegraphics[width = 8cm]{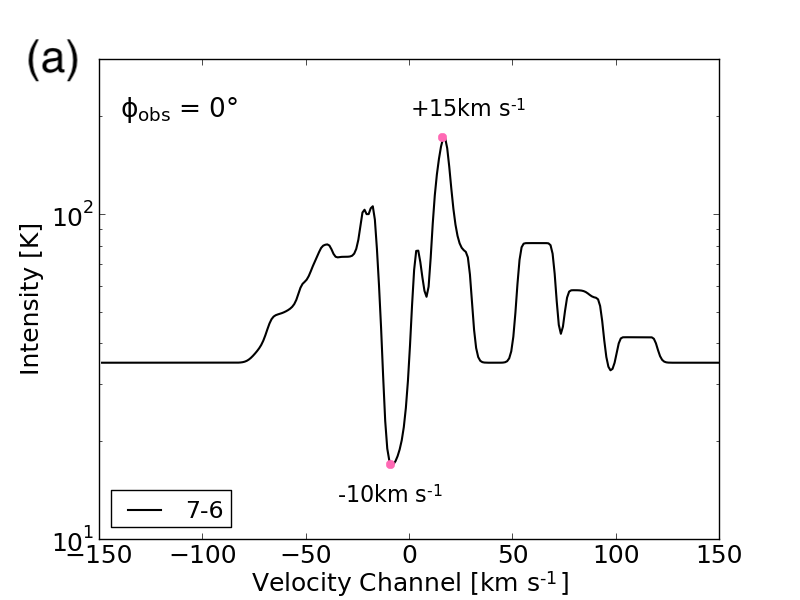}\\
\includegraphics[width = 7cm]{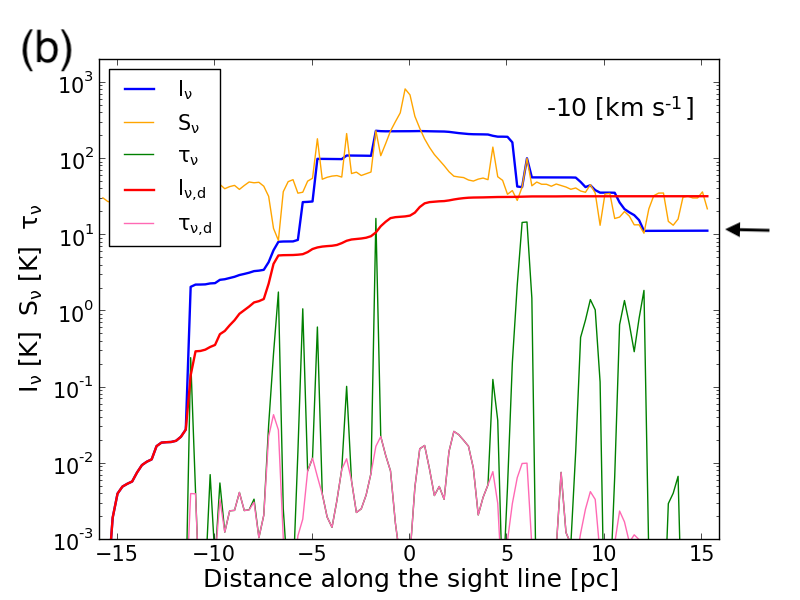}
\includegraphics[width = 7cm]{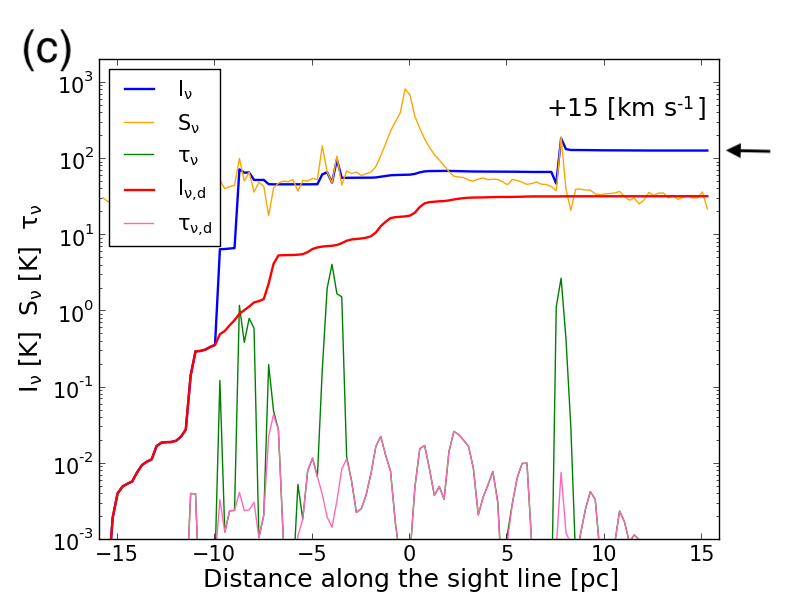} \\
\includegraphics[width = 10cm]{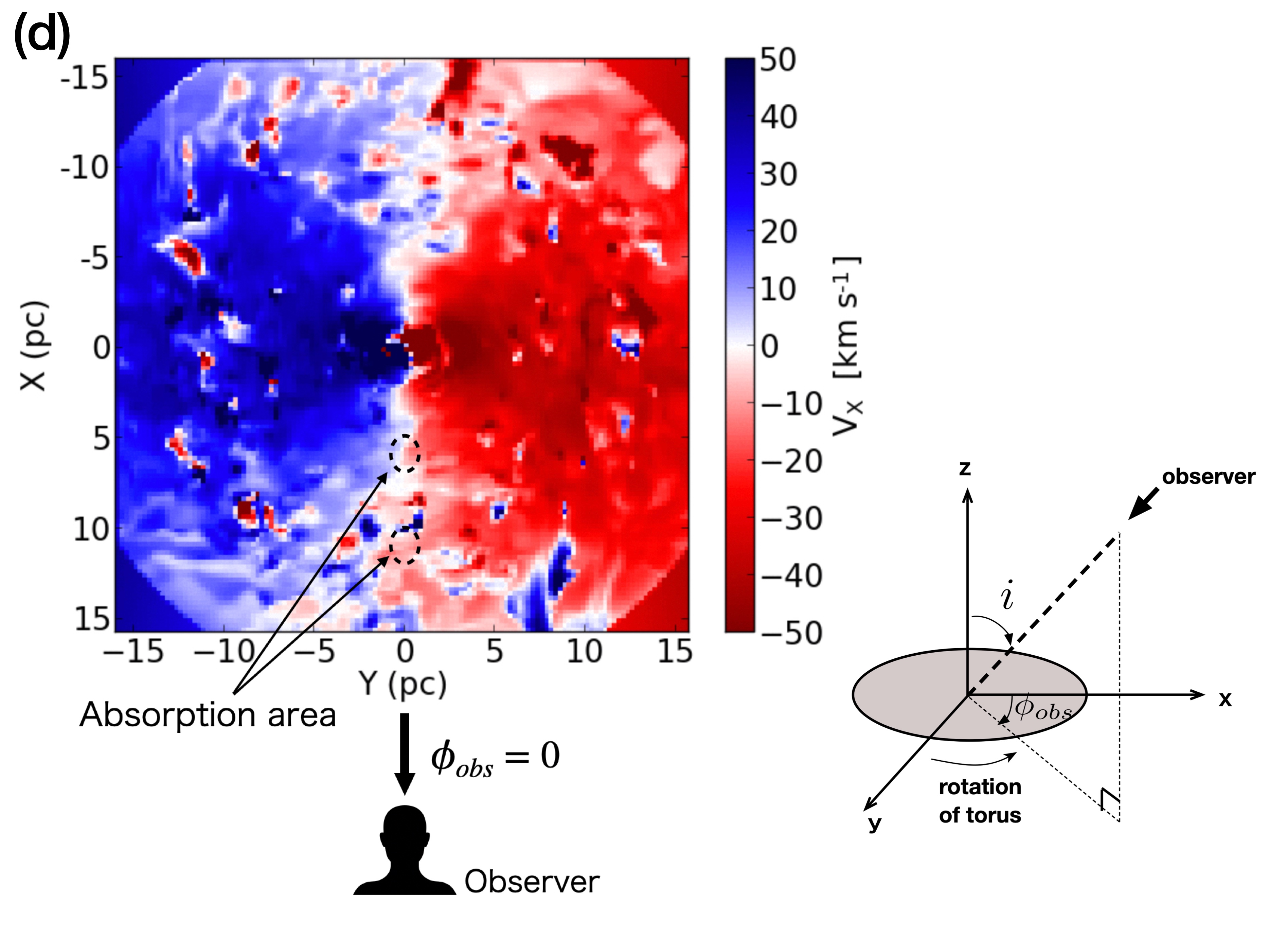}
\caption{(a) CO(7--6) spectrum (not continuum subtracted) at 371.7 $\mu {\rm m}$ (806.7 GHz) for $i = 90^\circ$, $\phi_{obs} = 0^\circ$, and $v_{turb} = 5$ \kms. The spectrum has been averaged to 0.5 pc$\times $ 0.5 pc resolution toward the center.
{Note that $v > 0$ here means that the gas moves toward $x > 0$ along the $x$-axis (i.e., outward but is not \textit{red-shifted}).}
(b) Total intensity $I_\nu$ (blue line), intensity of dust continuum $I_{\nu, d}$ (red line), source function $S_\nu$ (yellow line), total optical depth $\tau_\nu$ (green line), and optical depth of dust emission $\tau_d$ (pink line) along the line of sight toward the
galactic center. This plot corresponds to the deep absorption feature at $v = - 10$ \kms. 
The arrow represent the location of the observer.
(c) Same as (b) but for $v = +15$ \kms .
(d) Distribution of $v_x$ (i.e., $x$-component of the velocity of gas in the input hydrodynamic model) on the $z=0$ plane. 
Here, the observer is located at $i = 90^\circ$ and $\phi_{obs} = 0$. }
\label{wada_fig: 10}
\end{figure}

Conversely, for lower-$J$ transitions, such as CO(1--0) and CO(3--2), the intensity is positive throughout the velocity range. 
Independent emissions are also observed at $ |v| \gtrsim100 $ \kms, which correspond to high-velocity components with non-circular 
motions near the center ($ r < 5$ pc) (see Fig. \ref{wada_fig: 10}).

\begin{figure}[h]
\centering
\includegraphics[width = 8cm]{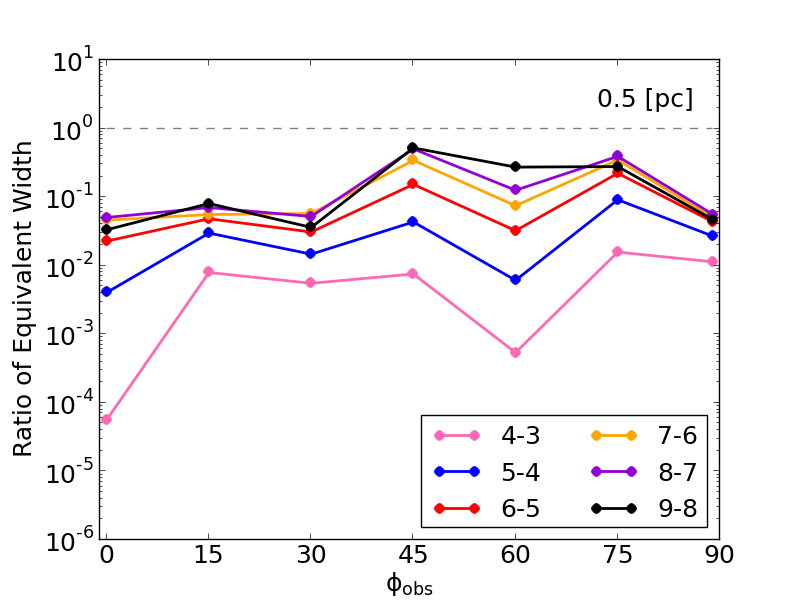}

\caption{Ratio of the equivalent width of an emission line to that of an absorption line ($f_{EW}$ given by Eq. (\ref{eq:ew})) 
as a function of the azimuthal angle ($\phi_{obs}$) in the edge-on view for a ``beam'' size of 0.5 pc. {Note that for $J=3-2$ and lower transitions, $f_{EW} \ll 10^{-4} $, which implies that the emission dominates for these transitions.} }
\label{wada_fig: 11}
\end{figure}

{Note that in Figs. \ref{wada_fig: 3} and \ref{wada_fig: 5}, we fixed the azimuthal angle of the
observer ($\phi_{obs}$, see Fig. \ref{wada_fig: 10}(d) ) to $50^\circ$. 
However, the input model used in this work, that is, the radiation-driven fountain model \citep{wada2016}, emplys a fully 3D calculation that does not assume axisymmetry. Moreover, the structure of the ISM is not necessarily homogeneous in the azimuthal direction, 
as also seen in Figs. \ref{wada_fig: map2} and \ref{wada_fig: 2b}(a).
Therefore, we expect the spectra to depend on $\phi_{obs}$ for a given inclination angle.
To quantify this, in Fig. \ref{wada_fig: 11}, we plot $f_{EW}$} as a function of the azimuthal angle for CO(1--0), ..., CO(9--8) at $i = 90^\circ$, where $f_{EW}$ is
the ratio of the equivalent width of an emission line to that of an absorption line against the continuum, 
defined as
\bea
f_{EW} \equiv \frac{ \int (1- I_\nu /I_{\nu, d} )  \, dv }{\int (I_\nu /I_{\nu, d} - 1) \, dv}. 
\label{eq:ew}
\eea
If $f_{EW} < 0$, then the spectral features are dominated by emission in the velocity range $v = -150$ to $+150$ \kms.
A larger value of $f_{EW}$ implies stronger absorption features, as seen in
Fig. \ref{wada_fig: 5} for a beam size of 0.5 pc.
{
Figure \ref{wada_fig: 11} shows that although the emission dominates for all azimuthal angles, there is more absorption at higher-$J$ transitions, such as $J=7-6, 8-7$, and $9-8$.
For transitions with $J = 6-5$ and higher, we find that $f_{EW} \gtrsim 0.1$. 
We also observe that 
there are weak variations in $f_{EW}$ with $\phi_{obs}$, which reflects the non-uniform
structure of the density and velocity in the torus.
}

\subsection{Origin of the spectral features}
{One of the main advantages of our model is that we can use it to identify the formation mechanism of the
absorption (or emission) line features in the obtained spectra as well as their location along the line of sight. 
This information is important for
understanding the structures of the molecular tori when interpreting spectra in future observations. To demonstrate this, in
Fig. \ref{wada_fig: 10}, we compare 
the spectrum, change in intensity along the line of sight and 
velocity field of CO(7--6) obtained using our input model.
Figure \ref{wada_fig: 10}(a) shows the CO(7--6) spectrum for $i = 90^\circ$ and  $\phi_{obs} = 0^\circ$. 
The spectrum has several positive and negative peaks\footnote{Note that in Fig. \ref{wada_fig: 10}, 
$v_x$ is observed along $\phi_{obs} = 0^\circ$ (i.e., $y = 0$); hence, positive velocities imply that the gas moves $\textit{outward}$ (i.e., blue-shifted), and negative velocities imply that the gas moves $\textit{inward}$ (i.e., red-shifted).}, but 
we only focused on the two prominent features at $v = + 15$ \kms and $v = -10$ \kms.
} 

In Figs. \ref{wada_fig: 10}(b) and (c), 
the observer is located on the extreme right (i.e., far beyond $+15$ pc).
Now, if the observed intensity $I_\nu$ (blue solid line) is less than the intensity of dust emission $I_{\nu, d}$ (red line), then it appears as an
absorption feature, else it appears as an emission feature. Fig. \ref{wada_fig: 10}(b) shows that the strong absorption feature at $-10$ \kms appears between 10 and 16 pc, which corresponds to the
outer region of the torus. 
The emission line feature at $+15$ \kms is mainly caused by the dense region with $\tau_\nu > 1$ at $r = 8$ pc.

These spectral features also reflect the internal velocity field of the torus. In Fig. \ref{wada_fig: 10}(d),
 the line-of-sight velocity ($v_x$) field is plotted
for an observer located at $y=0$ and $x = +\infty$. The absorption feature at $v = -10$ \kms corresponds to the 
\textit{inflow} component seen in red at $r \sim 10$ pc. This inflow component can also be observed on the opposite side (i.e., $x = -10$ pc and $y = 0$). 
The two strong absorption features at $v_x = -10$ \kms seen in Fig.  \ref{wada_fig: 10}(b) are indicated by dashed circles in Fig. \ref{wada_fig: 10}(d).
The emission line at $v = +15$ \kms seen in Fig. \ref{wada_fig: 10}(a) originates from one of the clumpy ``blue" regions along $y =0$. 
The clumpy regions around $r \sim 10$ pc observed in the velocity map are caused by supernova feedback
in the radiation-driven fountain model \citep{wada2012, wada2016}. 
This feedback makes the torus geometrically thick, as seen in Fig. \ref{wada_fig: 2}(a),
and hence, some of the emission features with high velocities in the CO spectra (see Fig. \ref{wada_fig: 5})
depend on the observer's azimuthal angle $\phi_{obs}$. 
This is also the reason for the weak dependence of $f_{EW}$ on $\phi_{obs}$ (see Fig. \ref{wada_fig: 11}). 

%
\section{Discussion and implications for future observations}
%
{
In recent ALMA observations, the molecular tori in some nearby Seyfert galaxies have been spatially resolved \citep{garciaburillo2014, garciaburillo2016, imanishi2016, imanishi2020, izumi2018, combes2019}.
According to the unified model of AGNs, the tori in type-2 AGNs should be close to edge-on; however, their inclination angles are not precisely determined.
In W16, the infrared SED was estimated using the same hydrodynamic model as in this work, and it was
suggested that the inclination angle of the torus should be larger than $i = 75^\circ$ in the Circinus galaxy
to explain the absorption feature at  $9.7~\mu$m in the observed SED.
This is consistent with the cone-like structure of the NLR observed in the Circinus galaxy (Paper III).
In this study, we investigated the dependence of the absorption features owing to the rotational transitions of CO on the inclination angle of the torus.}

\begin{figure}[h]
\centering
\includegraphics[width = 7cm]{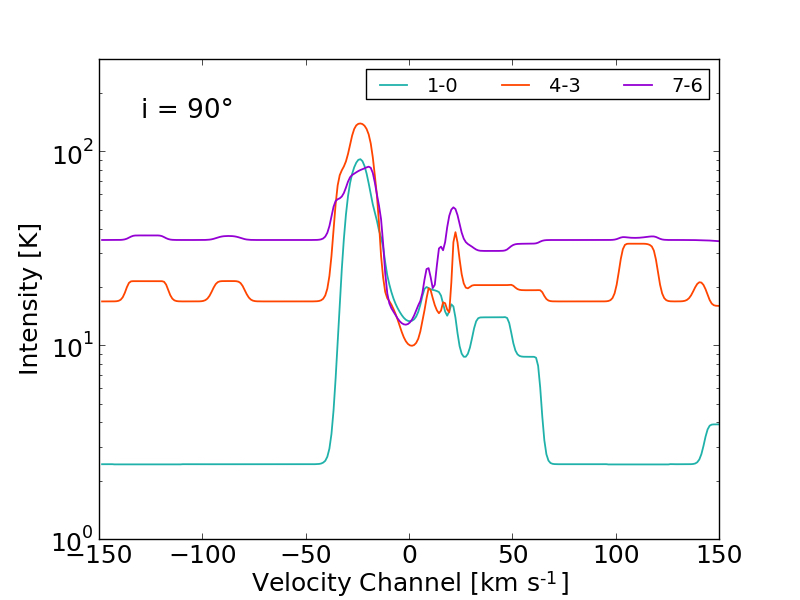}
\includegraphics[width = 7cm]{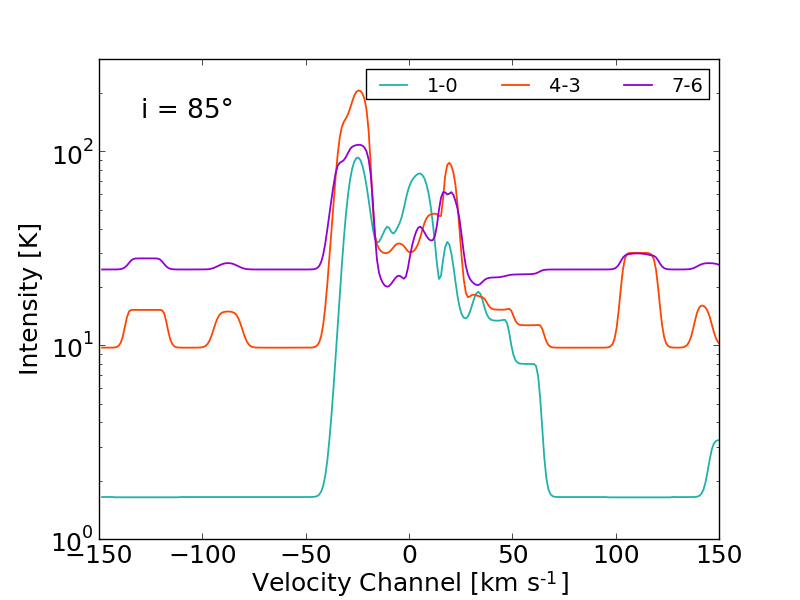}
\includegraphics[width = 7cm]{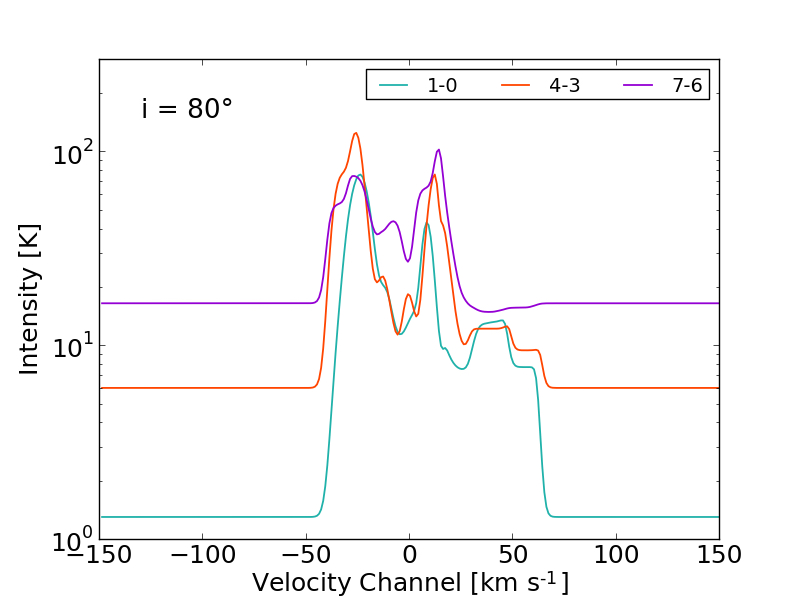}
\includegraphics[width = 7cm]{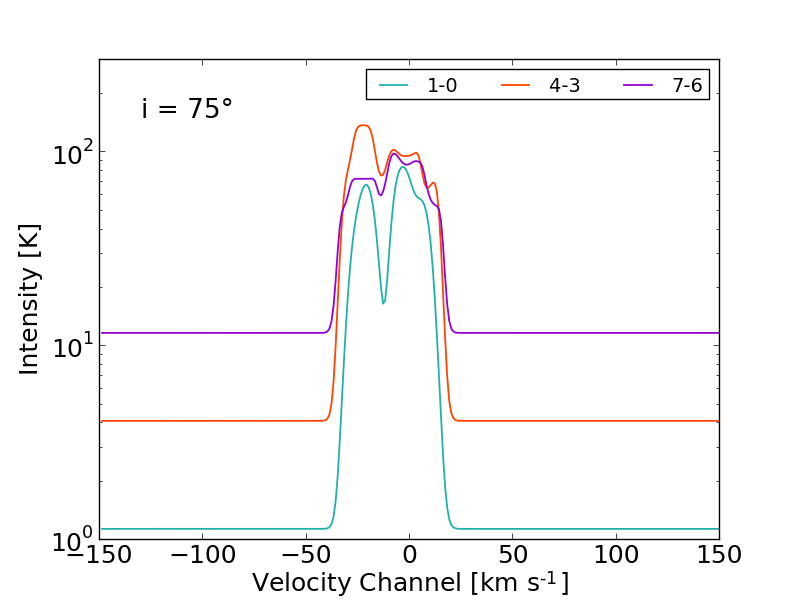}
\includegraphics[width = 7cm]{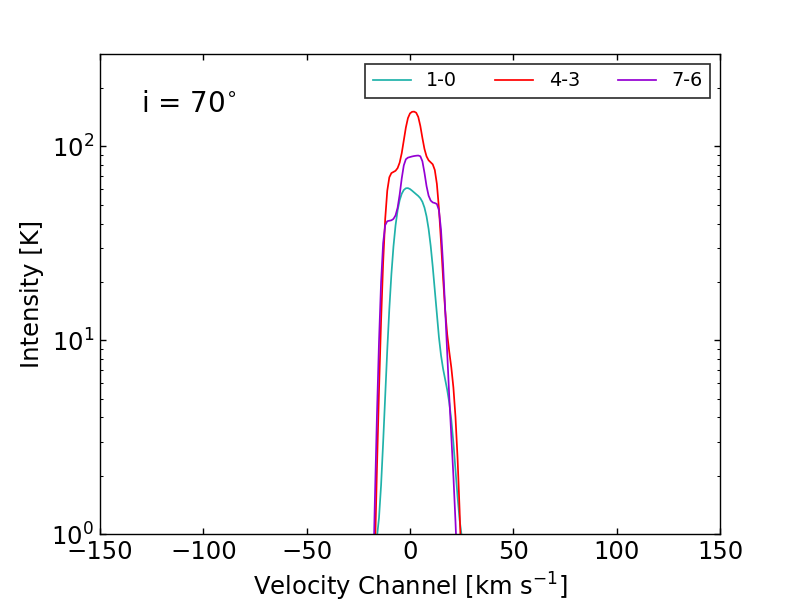}
\includegraphics[width = 7cm]{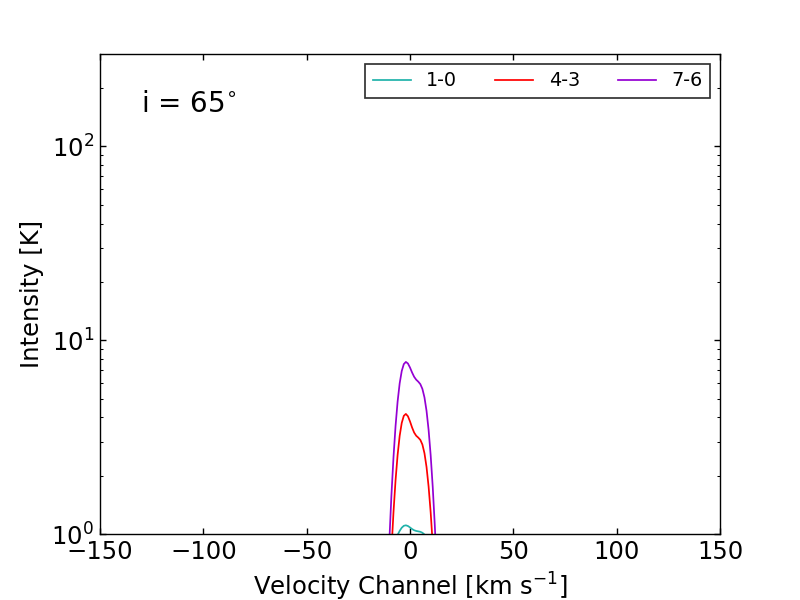}
\caption{{Dependence of the CO spectra ($J=7-6, 4-3$ and $1-0$ toward the AGN position on the inclination angle of the torus. The continuum unsubtracted spectra are plotted for $i = 90^\circ, 85^\circ, 80^\circ$, 75$^\circ$, 70$^\circ$, and 65$^\circ$. 
Here $\phi_{obs} = 50^\circ$, 
beam size of 0.5 pc, and $v_{turb} = 5$ \kms are assumed.}}
\label{wada_fig: 12}
\end{figure}

Fig. \ref{wada_fig: 12} shows the spectra of three transitions, CO(1--0), (4--3), and (7--6), for {six} different viewing angles
toward the nucleus. We observe that 
the deep absorption feature at $v = -10$ \kms in the CO(4--3) and (7--6) spectra becomes shallower for $i = 85^\circ$ and $80^\circ$; whereas no absorption features are observed for $i  = 75^\circ$, { or smaller}. 
{For the smaller viewing angles (i.e., close to face-on),  there is almost no dense molecular gas along the line of sight (see Fig. 
\ref{wada_fig: 2}), 
if the beam size is small, therefore we observe weak emissions even for CO (1-0).}
{It should be noted that the spectra also depend weakly on the azimuthal angle, because of  the internal inhomogeneous structure of the torus (see Fig. \ref{wada_fig: 11}). However, we find that the absorption features are more sensitive to the
inclination angle. {For example, for CO(7--6), absorption features against the continuum
are detected for five azimuthal directions out of nine for $0 <\phi_{obs} < \pi/2$ and  $i \ge 85^\circ$, } whereas no absorption features are observed against the continuum for $i \le 80^\circ$.
In other words, to detect absorption features in the continuum of the CO rotational transition lines, 
it is necessary (but not sufficient) for the torus to be close to edge-on, that is, $i \gtrsim 85^\circ$.
}

\begin{figure}[h]
\centering
\includegraphics[width = 8cm]{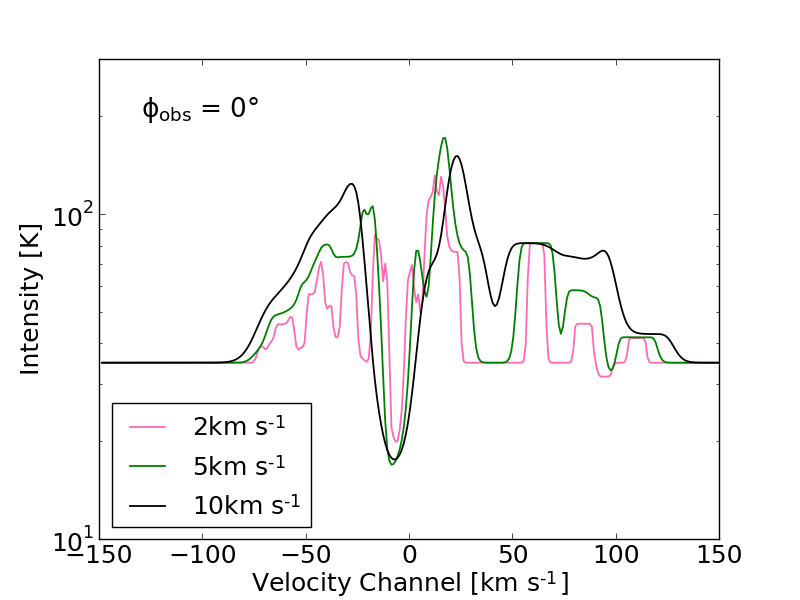}  
\includegraphics[width = 8cm]{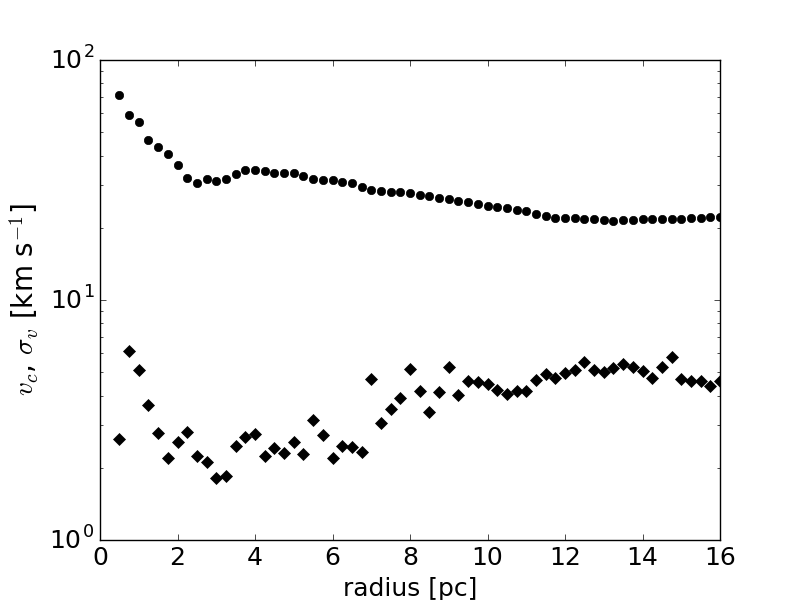}  

\caption{(top) Dependence of the CO(7--6)  spectra ({continuum unsubtracted}) on $v_{turb}$, for $i = 90^\circ$, $\phi_{obs} = 0^\circ$, and beam size of 0.5 pc.
(bottom) Radial distributions of the circular velocity ($v_c$) and the velocity dispersion ($\sigma_v$) for $\Delta r = 0.25$ pc.}

\label{wada_fig: 13}
\end{figure}

In the results presented in \S 3, we assumed that the internal turbulent velocity or ``micro"-turbulence 
in each grid cell of size (0.25 pc)$^3$ was 5 \kms. In Fig. \ref{wada_fig: 13}, we plot the CO(7--6) spectra for different turbulent velocities, namely,  $v_{turb} = 2, 5, $ and 10 \kms. The deepest absorption line at $v = -10$ \kms appears for all three cases, but 
the line width decreases with the turbulent velocity.
Moreover, the broad emission line features seen for $v_{turb} = 10$ \kms are split into multiple emission and absorption lines for $v_{turb} = 2$ \kms. 

{
 If the continuum absorption is observed in nearby type-2 Seyfert galaxies (e.g., the Circinus galaxy) by future high-resolution observations, we could obtain 
 some information about the internal velocity structures of the molecular torus.
 Our results suggest that the width of the continuum absorption depends on the assumed turbulence velocity below the spatial scale of 0.25 pc; it is $\sim 9$ \kms for $\sigma_v = 5$ \kms at FWHM, and $\sim  5$ \kms for $\sigma_v =$ 2.5 \kms.
As shown in \S 3, the continuum absorption is mainly generated from the gas around r = 10-15 pc (Fig.  \ref{wada_fig: 10}), 
where the velocity dispersion of the gas in the input model is $\sim 5$ \kms;  whereas the rotational velocity at the radius is
30 \kms  (Fig. \ref{wada_fig: 13}(bottom)).   Therefore, the line width of the continuum absorption seen in the spectra
reflects the internal velocity dispersion of the ISM 
rather than the velocity shear.
}

{
On the other hand, the self-absorption features seen in the spectra are difficult to be understood in principle because it cannot be distinguished from multiple emission lines.
However, Fig. \ref{wada_fig: 13} shows that 
the emission lines for  $v_{turb} =$ 2 \kms and 10 \kms are clearly different. 
In the former case, the emission features around the continuum absorption are separated into several components. Thus, the 
structures of the emission features may also reflect to the internal velocity field on a sub-pc scale. }


\section{Conclusions}

{Absorption features owing to silicates and CO rotation-vibration transitions are powerful tools for understanding the physical conditions and structure of the dense ISM
in AGNs, such as ULIRGs \citep[see e.g.,][]{spoon2004, shirahata2013, shirahata2017, baba2018}.
However, absorption features of molecular lines in the millimeter and submillimeter wavelength range
have never been observed in Seyfert galaxies \citep{okuda2013}, where
the dust in the molecular tori is heated by the nuclear radiation.
Although the dust temperature is not high (i.e., $<$ 100 K), it could still lead to the formation of a background source. 
Therefore, it is important to understand the conditions necessary to observe CO absorption features against the dust continuum.}

{
To address this problem theoretically, in this work,
we used a realistic 3D physical model 
that includes both the dust component and molecular gas distribution. Specifically, we used
the radiation-driven fountain model \citep{wada2012, wada2016}, 
  from which we calculated both the background dust continuum 
  and molecular line intensities.
  The temperature of the dust associated with the ISM was self-consistently determined 
  from the radiation field of the AGN.
The hydrodynamic model employed
consistently explains the multi-wavelength observations of the Circinus galaxy, as discussed in
the previous papers of this series (Papers I--III).
}

{In contrast to Papers I and II, 
in this study, we conducted new three-dimensional radiative transfer calculations 
including both the continuum emission owing to the AGN-heated dust 
and the emission owing to the rotational transition lines of CO. 
We investigated the spectral properties of several CO rotational lines ($J=1-0, 2-1, ...$) and calculated
the dust thermal emission
 using the three-dimensional radiative transfer code } \verb+RADMC-3D+ \citep{dullemond2012}. 

 {Our main findings can be summarized as follows:  }
 
{(1) We 
 studied the absorption features against 
the continuum originating from the AGN-heated dust of several tens of kelvin in the radiation-driven fountain model.
We found that the continuum-subtracted channel maps of $J = 4-3$ and higher transitions show 
absorption regions along the disk mid-plane. 
For lower-$J$ transitions (e.g., $J=1-0$ and $2-1$), emission features dominate, but self-absorption features 
exist at low velocities ($ \le \pm 50$ \kms).
The spectra consist of multiple absorption and emission features reflecting the complicated structure of
the density and velocity in the torus.
The deepest absorption feature is caused by the infalling gas on the near-side of the torus 
between $r \sim10$ and 15 pc, which is located in front of the heated dust within $r \simeq 5$ pc.
}

{
(2) We found the conditions necessary to detect CO absorption features against the continuum, which are: (a) the transitions should be at
$J = 4-3$ or higher, (b) the spatial resolution should be high, in the range 0.5--1.0 pc, and c) the tori should be observed
nearly edge-on (i.e., $i \gtrsim 85^\circ$ for CO(7--6)). Conditions (a) and (b) can be achieved by ALMA 
for nearby AGNs.}   

{
(3) We also found the  correlation between the absorption features and 
the internal spatial and kinematic structure of the obscuring material, 
such as the bulk motion (e.g., inflow toward the center) and turbulent motion. 
Moreover, the internal turbulent velocity ($v_{turb}$) in each grid cell (i.e., (0.25 pc)$^3$ in the present work) was found to
affect the 
width of the {continuum} absorption lines, which was established using $v_{turb} = 2, 5$, and 10 \kms.}

{
The results presented in this work imply that appropriate conditions are required for observing CO absorption features owing to 
rotational transitions in the submillimeter range in nearby Seyfert galaxies. Nevertheless, our model can provide
new information about the internal structure of the molecular tori in nearby AGNs
when combined with ALMA observations, for example, of CO(7--6) at Band 10.
}

\acknowledgments
The authors thank the anonymous referee for his/her many comments and suggestions to improve this paper. 
We also thank J. P. Perez-Beaupuits for his valuable comments on the XDR chemistry.
Numerical computations were performed on a Cray XC50 supercomputer at the Center for Computational Astrophysics, National Astronomical Observatory of Japan. This work was supported by the Japan Society for the Promotion of Science (JSPS) KAKENHI Grant Numbers 16H03959 and 21H04496.
T.I. and S.B. were supported by JSPS KAKENHI Grant Numbers JP20K14531 and JP19J00892, respectively.
Y.K. was supported by the NAOJ ALMA Scientific Research Grant (No. 2020-14A).

\section*{Appendix}

\begin{figure}[h]
\centering
\includegraphics[width = 6.5cm]{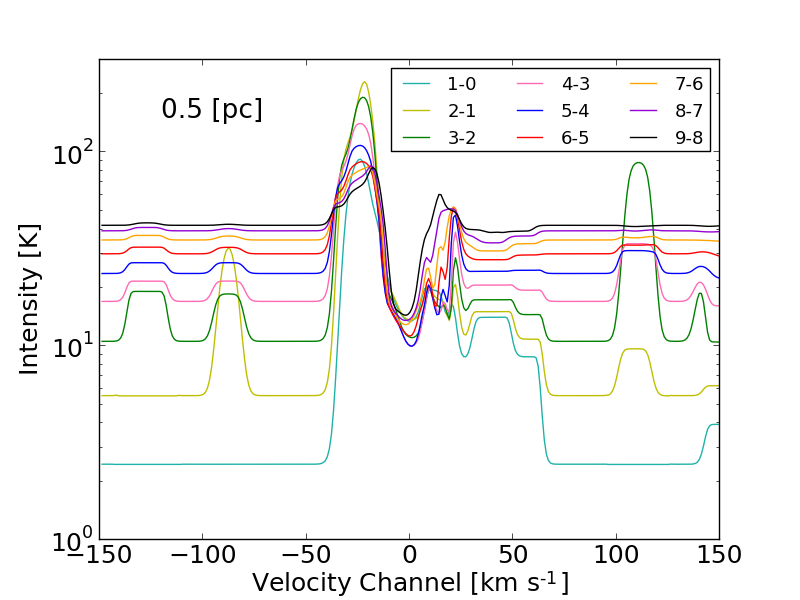} 
\includegraphics[width = 6.5cm]{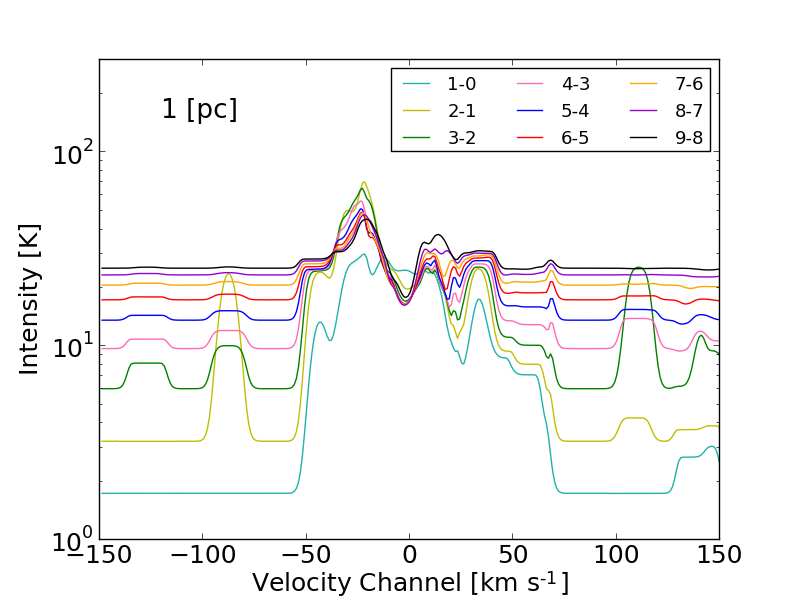} \\
\includegraphics[width = 6.5cm]{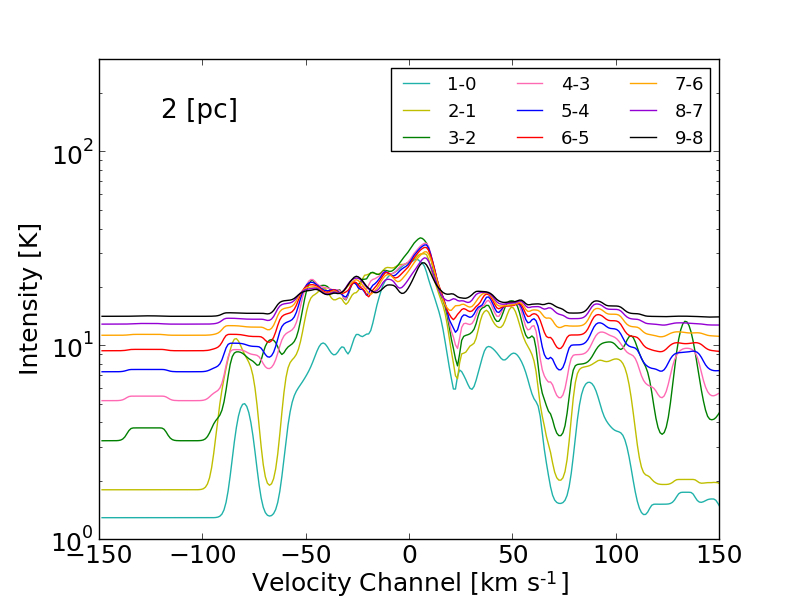}
\includegraphics[width = 6.5cm]{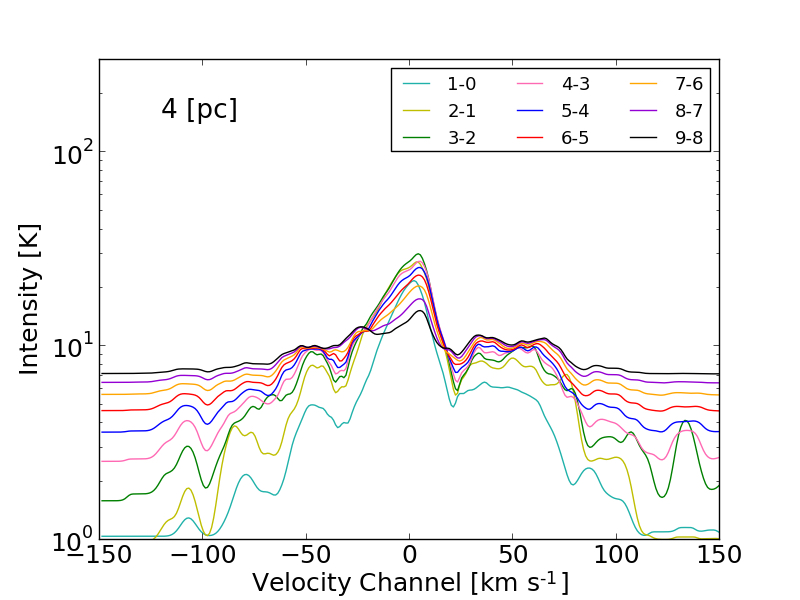}
\caption{CO rotational transition spectra  (continuum unsubtracted) toward the center without subtracting the dust continuum, for $i = 90^\circ$ (edge-on), $\phi_{obs} = 0^\circ$, and $v_{turb} = 5$ \kms. The beam sizes from top to bottom are 0.5, 1, 2, and 4 pc.  }
\label{wada_fig: appendix}
\end{figure}

Figure \ref{wada_fig: appendix} 
shows the same spectra as in Fig. \ref{wada_fig: 5} 
but without subtracting the dust continuum.  
These raw spectra show that 
the continuum levels are higher for higher $J$ transitions.
The deep absorption features in the range $-10$ to $0$ \kms for beam sizes 0.5 and 1 pc are common for
most transitions; however, they transform into self-absorption features if the continuum level is extremely high.

\newpage

\end{document}